\documentclass[final,5p,times,twocolumn]{elsarticle}

\usepackage{amsmath}
\usepackage{amssymb}
\usepackage{epstopdf}
\usepackage{graphicx}
\usepackage{etoolbox}
\apptocmd{\sloppy}{\hbadness 10000\relax}{}{}

\journal{Physics Letters B}

\begin{document}

\begin{frontmatter}



\title{A unified nucleosynthetic site  for the production of heavy isotopes and p-nuclei}


\author{Amir Ouyed, Rachid Ouyed, and Denis Leahy}

\address{Department of Physics and Astronomy, University of Calgary,
2500 University Drive NW, Calgary, Alberta, T2N 1N4 Canada}

\begin{abstract}
Current r-process models  under-produce $A<130$ nuclei. P-process models either underproduce $A<100$ and  $150<A<165$ p-nuclei, or have other limitations. We argue that the puzzles in these two  mechanisms can be solved by heavy-ion spallation. Spallation happens in the explosive phase transition from a neutron star to a quark star - the Quark Nova (QN). The QN triggers the r-process which creates abundant $A>130$ nuclei, and spallation fragments these isotopes into $A<130$ nuclei and all 35 p-nuclei. Our model is universal in relation to a star's age, metallicity, and chemistry.

\end{abstract}


\end{frontmatter}



\section{Introduction}

	Currently there are three processes that are thought to dominate the nucleosynthesis of isotopes heavier than  $^{56}$Fe: the r-process, the s-process, and the p-process.  Roughly,  the r-process dominates the production of the vast majority of $A>90$ isotopes, while the s-process produces most $56<A<90$ isotopes. Yet, the p-process still plays an important role: the p-nuclides, which are 35 in total,  cannot be reproduced by neutron-capture processes and require the p-process \citep{rauscher2002nucleosynthesis}. For the p-process, researchers postulate  the photo-disintegration ($\gamma$-process) of seed nuclei as the dominant mechanism for the production of p-nuclei \citep{rauscher2002nucleosynthesis}. Unfortunately, the r-process is still poorly understood \citep{arnould2012r}, and  p-process models face many challenges \citep{rauscher2013constraining}. 

	Historically, the main puzzle of the r-process has been its astrophysical site. Traditionally, the prime candidate for the r-process site has been Type II Supernova (SNII)\citep{arnould2012r}. However current simulations can only detonate underpowered, mid mass-range supernovae \citep{RevModPhys.85.245}. Furthermore, current SNII models cannot generate enough neutrons nor entropy to trigger the r-process \citep{roberts2010integrated}. The second most popular candidate is the the Neutron Star Merger (NSM). Yet, NSMs happen at a relatively late stage of the universe's evolution, which  disagrees with the r-process enrichment of old, metal-poor stars \citep{jaikumar2007nucleosynthesis}. Furthermore, NSM's mass yield for elements lighter than $A \sim 140$ is too low in comparison with solar abundances \citep{korobkin2012astrophysical}.

	The most glaring issue with current p-process models is the underproduction of isotopes in  certain mass-ranges in comparison to solar abundances. The prime candidate for the p-process, the SNII, under-produces p-nuclei in the mass ranges   $150<A <165$, and  $A \le 100$ \citep{fuller1995neutrino,howard1991new,rayet1995p}. Researchers have suggested other candidates, such as Type Ia supernovae \citep{goriely2002he,howard1991new}, thermonuclear explosions on a neutron star's surface \citep{schatz2001end}, neutron-winds acting on hot-matter \citep{frohlich2006neutrino}, and accretion discs around compact objects \citep{fujimoto2003p}.  However,  no alternative to SNII has gathered a majority consensus. Furthermore, these models still lack consistent hydrodynamic and nucleosynthetic treatment \citep{rauscher2013constraining}. Another challenge  concerns the production of the very rare elements $^{138}$La and $^{180m}$Ta . The $\gamma$-process, which is thought of as the origin of most p-nuclei, cannot produce $^{138}$La and $^{180m}$Ta.   Researchers have suggested that these rare isotopes were produced by the interaction of neutrinos with stable nuclei ($\nu$-process) \citep{woosley1990nu,heger2005neutrino}. Nonetheless, this $\nu$-process for  $^{180m}$Ta has some uncertainties, because it is difficult to calculate   $^{180m}$Ta's final isomeric state \citep{mohr2007survival,belic2002photo}. Finally, the p-nuclei yield is sensible to details in a star's structure, including metallicity, and details in stellar evolution, which poses a problem when calculating the final, Galactic p-nuclei yields \citep{rauscher2013constraining}. 	
		
	Perhaps, the puzzles in both the p-process and r-process are closely related. We argue that the isotopic abundances that are attributed to the r-process and p-process are strongly affected by the presence of an alternate site - the Quark Nova (QN). The Quark-Nova is  the name given to the explosive phase transition from a neutron star to a quark-star \citep{ouyed2002quark}. 
	Although the QN model has not yet been used to explain p-nuclei, it has been applied to the aforementioned r-process puzzle. The hot, and neutron-dense matter ejected by the QN was suggested as an ideal site for the r-process\citep{ouyed2002quark}. Although, the QN was an excellent source of very heavy, $A>130$ isotopes, it strongly under-produced isotopes in the $A<130$ range, which effectively left the $90<A<130$ solar abundances still unexplained.  
	
	However, in this paper we consider the QN a site for heavy ion spallation. We argue that heavy-ion spallation  can successfully enhance isotopes in the $90<A<130$ mass range and thus, effectively complete the QN's r-process. Furthermore, this  $A>130$ yield also includes all 35 p-nuclei. 
	
	Spallation in the QN is triggered in the following way: If there is a short time delay between an SN II and the evolution of its neutron star into a QN, the relativistic, neutron-rich ejecta released by the QN will interact with the ejecta of the aforementioned SNII, triggering spallation reactions. This interaction between the QN's ejecta and the SNII's ejecta is referred as the dual-shock Quark-Nova (dsQN). The QN's ejecta is rich in  $A>130$ r-process isotopes, and these isotopes spall against the SNII's ejecta. In spallation, the $A>130$ r-process isotopes fragment into smaller daughter isotopes, which eventually turn into p-nuclei and other r-process nuclei. The masses of these spallation products covers the whole  $A<130$ range. Spallation in dsQN has been previously explored  \citep{ouyed2011}. However, not until now do we consider the fragmentation of $A>130$ r-process isotopes produced in the QN, which produces a much wider mass spectrum.

	\section{Model}\label{model}
	We use a similar model to \citep{ouyed2011}, but for nucleus-nucleus reactions. The reaction cross sections are calculated by using parameterized, partial cross sections for proton-nucleus reactions \citep{silberberg1998updated}, and scaling the proton-nucleus cross sections into nucleus-nucleus cross sections through the algorithm of \citep{tsao1993scaling}. We also make us of parameterized total nucleus-nucleus reaction cross sections as well \citep{sihver1993total}.  In order to visualize the model, we make use of particle accelerator analogies, wherein the experiment is divided into a beam, and a target. 
	
\textit{Initial Beam}: According to past studies \citep{keranen2005neutrino,ouyed2005fireballs},  an exploding neutron star can eject  its outer most layers with a mass  averaging $M_{QN}\sim 10^{-3} \text{ M}_{\odot}$. This neutron-rich ejecta becomes a site for r-process synthesis \citep{jaikumar2007nucleosynthesis}. The QN r-process generates a heavy isotope beam. We model the beam of heavy isotopes as a pulse with a mass of $\sim 10^{-5} \text{ M}_{\odot}$  since  the post-r-process QN  
	 ejecta is still rich in neutrons \cite{jaikumar2007nucleosynthesis}. We include the whole spectrum of isotopes produced in the QN r-process in the beam \citep{jaikumar2007nucleosynthesis} . The beam has an energy of $E \sim 10$ GeV/u,  which corresponds to a typical nucleon energy in the QN ejecta. 
	
\textit{Target}: The SNII's inner shell is modelled as the target. The inner shell is  a $^{56}$Ni  layer, with a mass  $0.1 \text{ M}_{\odot}$ of $^{56}$Ni. The shell is spherical, and expands with an inner radius of $R=v_{\text{sn}} \times t_{\text{delay}}$, where $v_{\text{sn}}=5000$ km/s is the speed of the SN ejecta, and $t_{delay}$ is the time delay between the first SNII detonation, and the subsequent, QN detonation.  The target's number density in the $^{56}$Ni  layer is approximately radially constant, at $n=N_{\text{Ni}}/(4 \pi R^2 \Delta R)$, where $N_{\text{Ni}}$ is the total number of $^{56}$Ni nuclei, and $\Delta R$ is the Ni layer's thickness.

\textit{Collision Statistics}: The probability that an inelastic collision between projectile specie $i$ and target specie $j$, produces fragment of specie $k$ is \\ \\ 
	
	\begin{equation} P \approx \frac{\sigma_{i \rightarrow k}(Z_i,A_i,Z_j,A_j, Z_k,A_k,E)}{\sigma_{\text{total},ij}(Z_i,A_i,Z_j,A_j)}\ , \end{equation}
	
	where $\sigma_{i \rightarrow k}$ is the partial reaction cross section for the fragmentation of projectile $i$ into fragment $k$. The total inelastic reaction cross section for projectile $i$ and target $j$, is  $\sigma_{\text{total},ij}$.
	 	
	When the projectile inelastically collides with a target nucleus, the number $N$ of nuclei $k$ synthesized in the collision is,
	
	\begin{equation}N_k \approx \frac{\sigma_{i \rightarrow k}}{\sigma_{\text{total},ij}} N_i \ . \end{equation}
	
	Where $ N_i$ is the number of projectiles $i$ before the collision. If the QN detonation happens with a time delay of a few days, then the SN ejecta would have much less entropy than the hot, ultra-relativistic QN r-process beam. Thus, in comparison to the beam, we approximate the SN ejecta to the lattice-like targets used in heavy ion collision experiments. Experiments for these lattice-like targets show that projectile fragment velocities remain roughly the same to the initial velocity \citep{hufner1985heavy}. Therefore we assume that projectile fragments conserve the same $E \sim 10$ GeV/u, and the target's fragments remain in their original place. Although the target (SN) ejecta has  $v_{\text{sn}}=5000$ km/s speed, we treat the target as static in comparison to the beam's ultra relativistic velocity.  The cross-sections are calculated in the projectile's rest frame.
	
	The inelastic collision events are modelled in the following way. We divide the target slab radially into virtual layers. Because the projectile beam is modelled as a pulse, all projectiles collide inelastically. against target nuclei simultaneously at each of the individual layers. When the beam collides against a layer, the projectile beam fragments into smaller projectiles. Thus, the beam's composition evolves at each collision, and therefore these smaller fragments collide in the subsequent virtual layers. We define an average,  free path $\overline{\lambda}$ as a radial separation between the layers,
	
	\begin{equation} \overline{\lambda}=\frac{1/n\sum_i N_i /{\sigma_{\text{total},i}}}{\sum_i N_i }\ , \end{equation}
	
	where $i$ is summed across all projectile species. The average, free path $\overline{\lambda}$ is calculated for each individual virtual layer. After the projectile beam collides with a layer, the average displacement of the beam becomes, 
		
	\begin{equation}x=\sum_q \overline{\lambda} _q \ ,\end{equation}
	
	where $q$ is summed across all virtual layers traversed by the beam. The simulation stops when the displacement becomes larger than the width of the slab, that is, when the $x \le \Delta R$ condition fails to be met.

\section{Results}

	\begin{figure}
 \centering
    \includegraphics[width=0.45\textwidth]{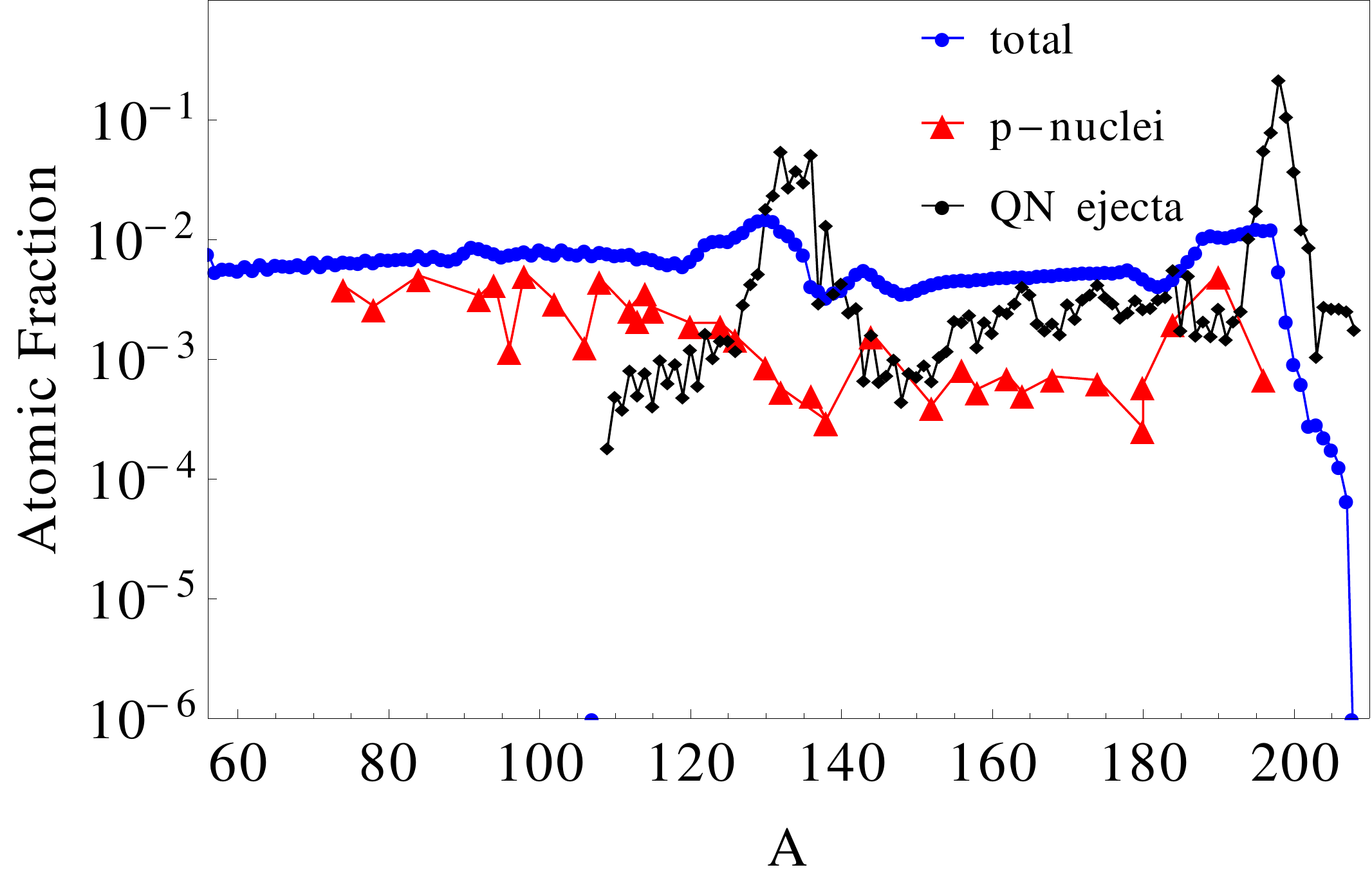} 
    \includegraphics[width=0.45\textwidth]{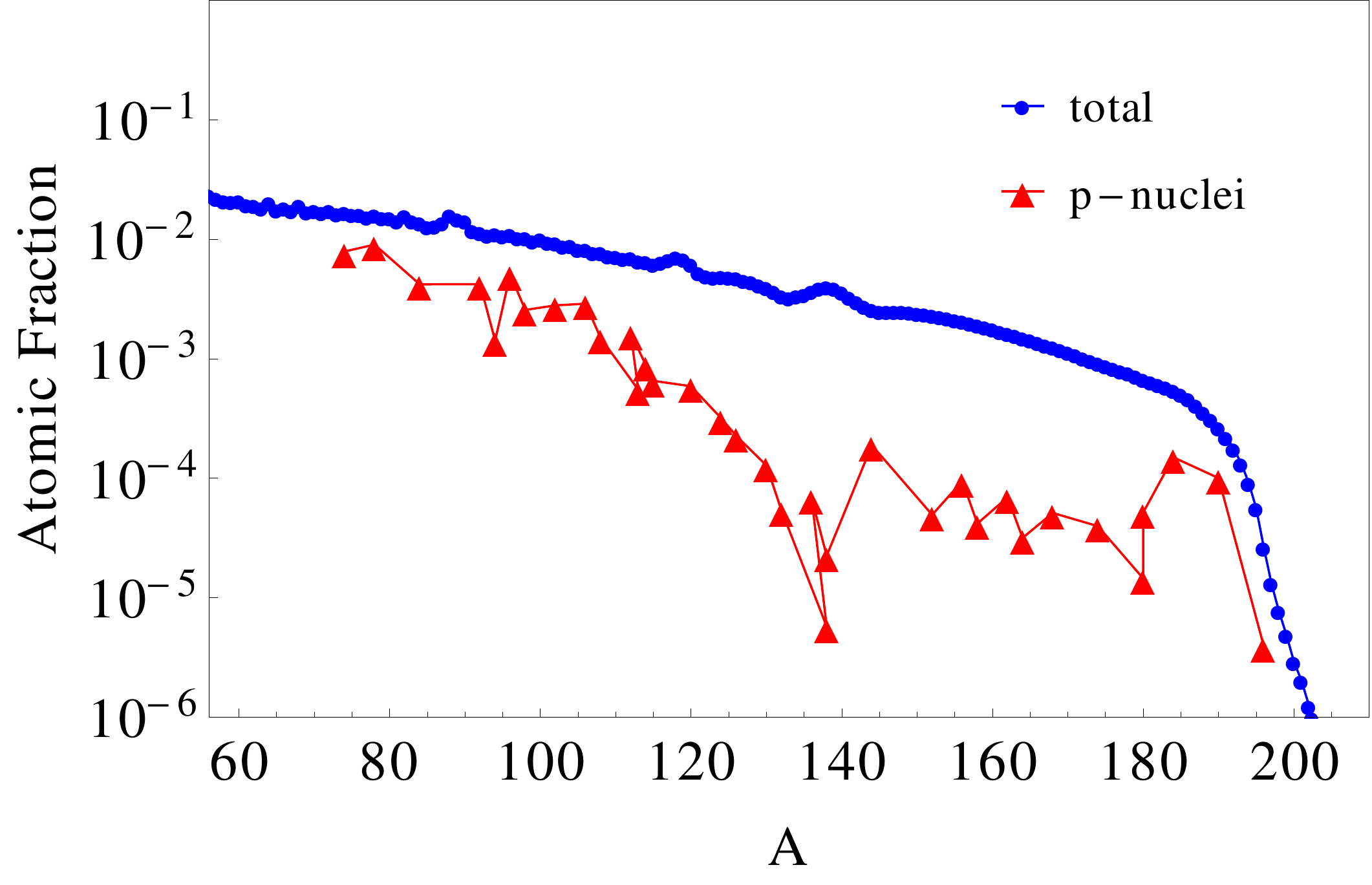} 
        \includegraphics[width=0.45\textwidth]{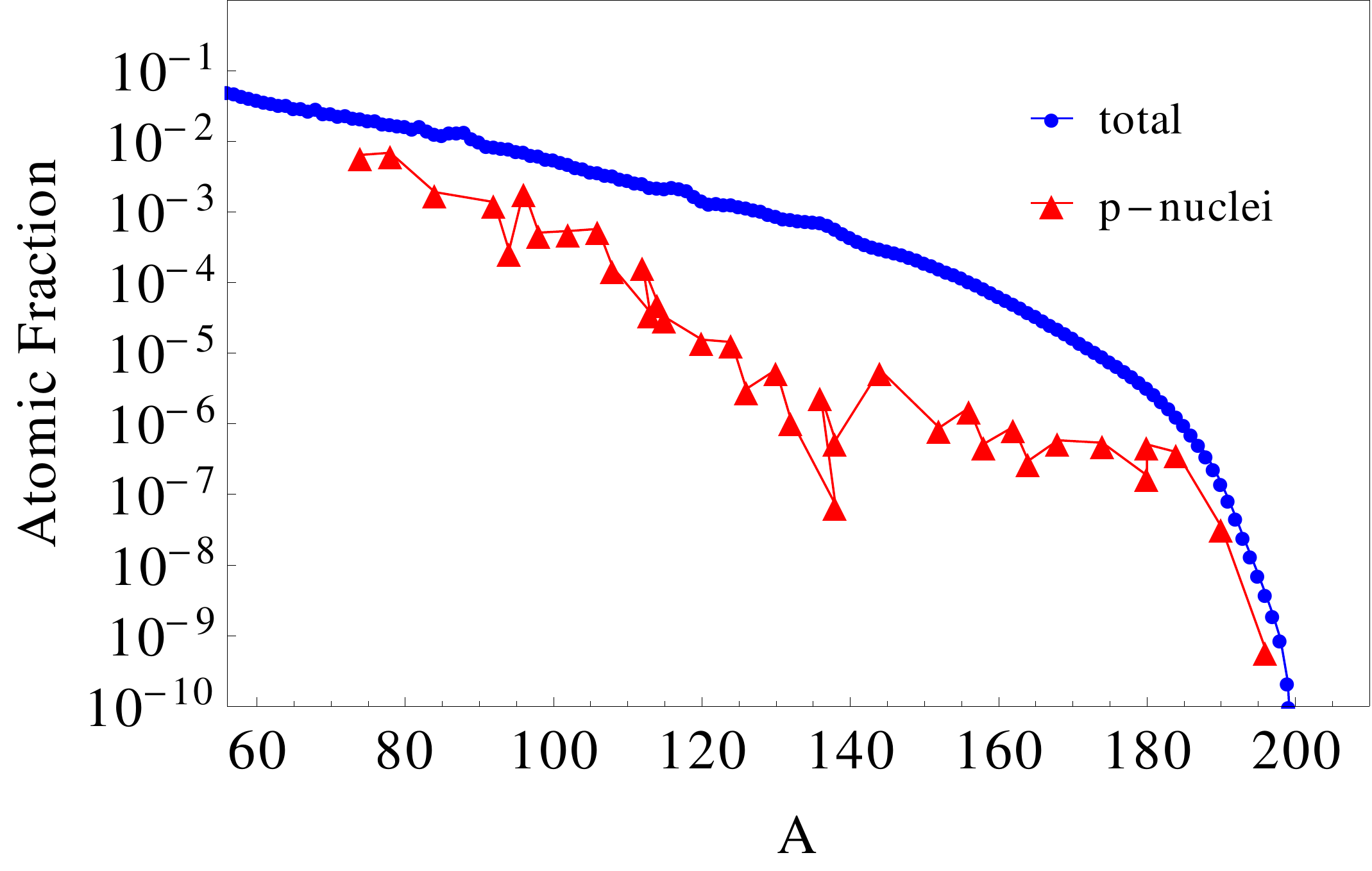}
  \caption{ \label{plots1}  Total  daughter $A>56 $ isotopes and p-nuclei yields produced by  the spallation of the QN r-process beam as a function of $A$. Blue dots are total abundances produced by spallation and red triangles are p-nuclei.  The plots are from top to bottom, $t_{\text{delay}} ({\rm days})=$ 5,3 and 2, respectively.  The initial, non-spalled QN r-process ejecta was graphed as black circles in the top plot for reference. Abundances  were normalized so that the sum of total abundances $A>56$ produced by spallation is unity.}
 
\end{figure}

\begin{figure}
 \centering
    \includegraphics[width=0.45\textwidth]{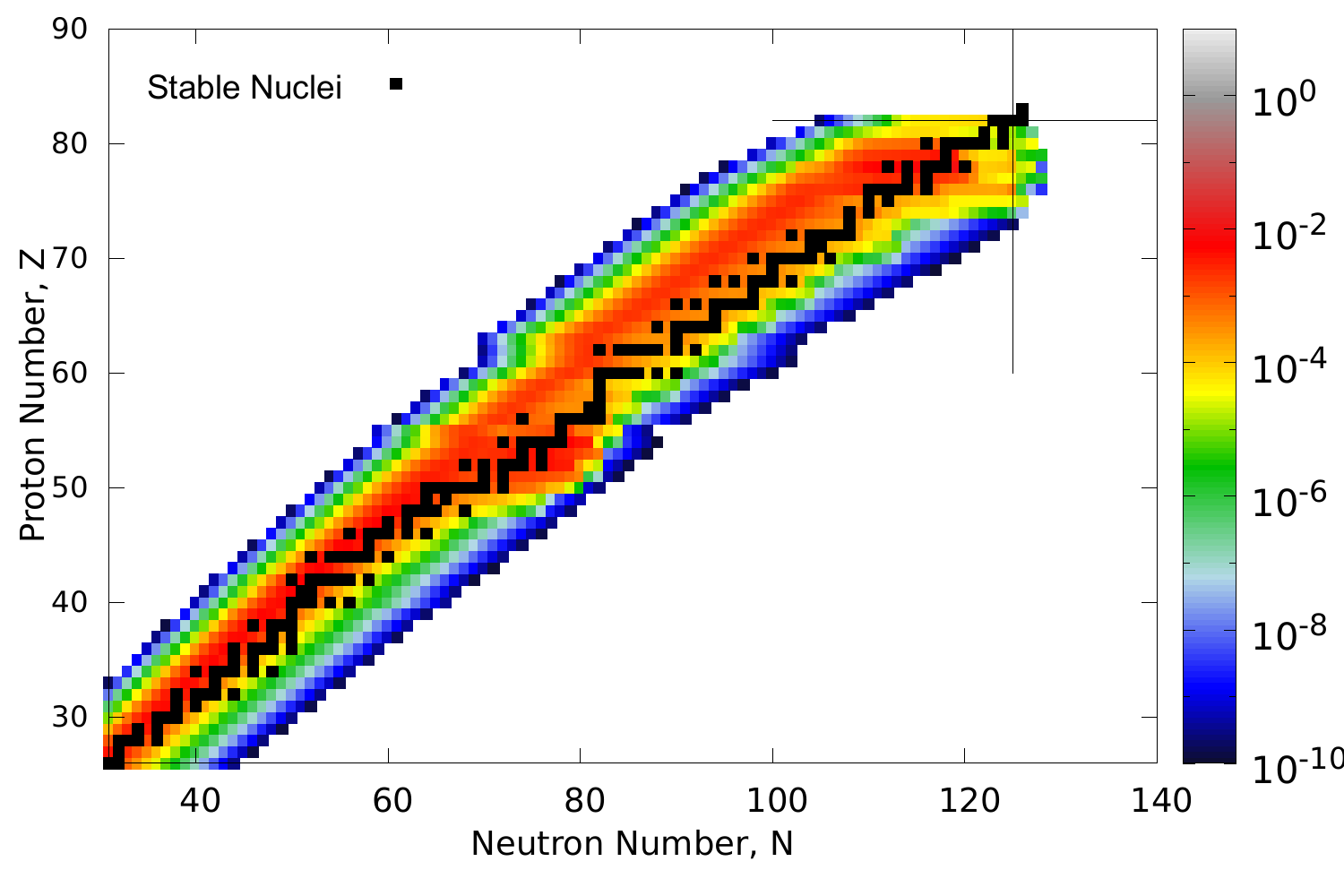} 
    \includegraphics[width=0.45\textwidth]{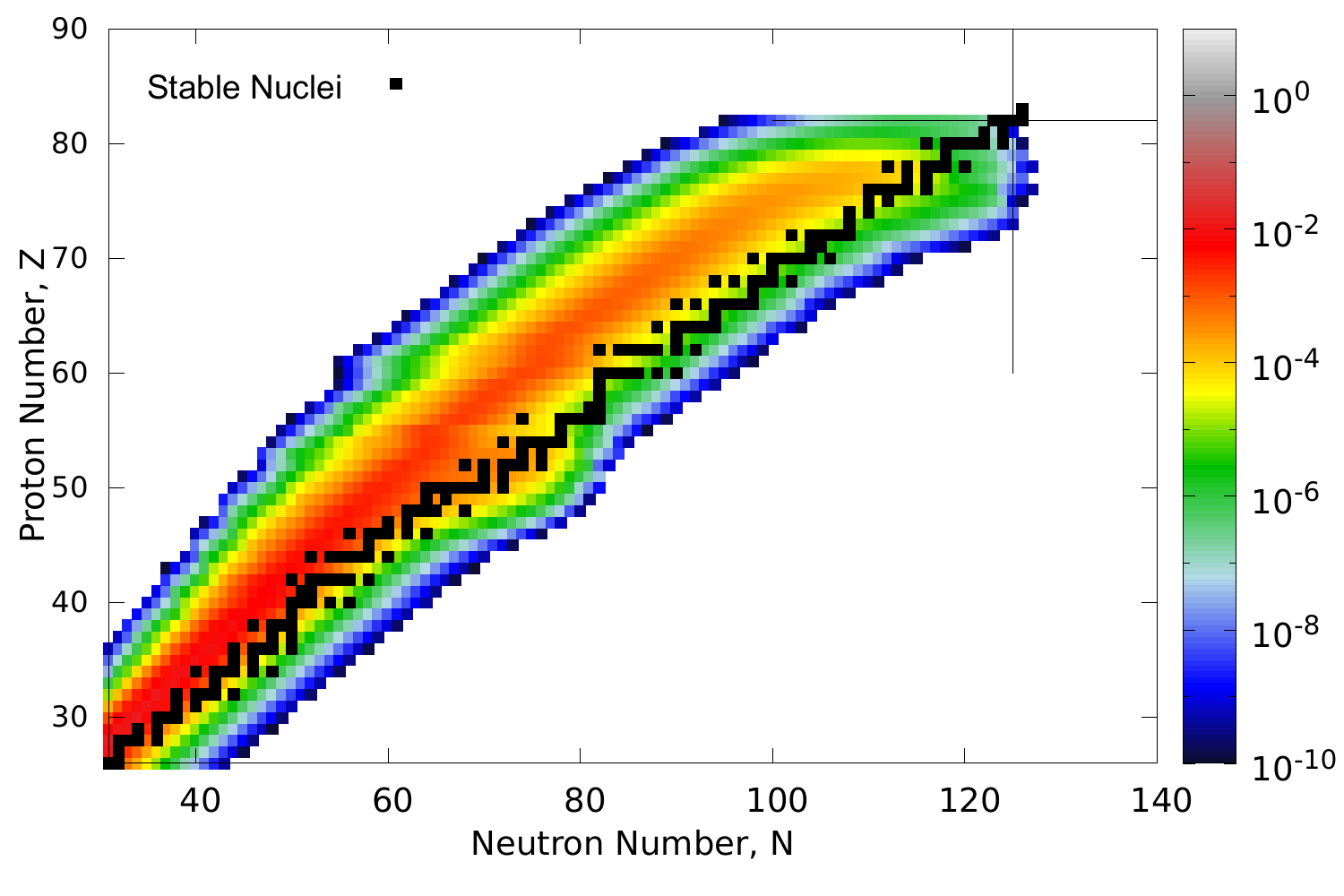} 
        \includegraphics[width=0.45\textwidth]{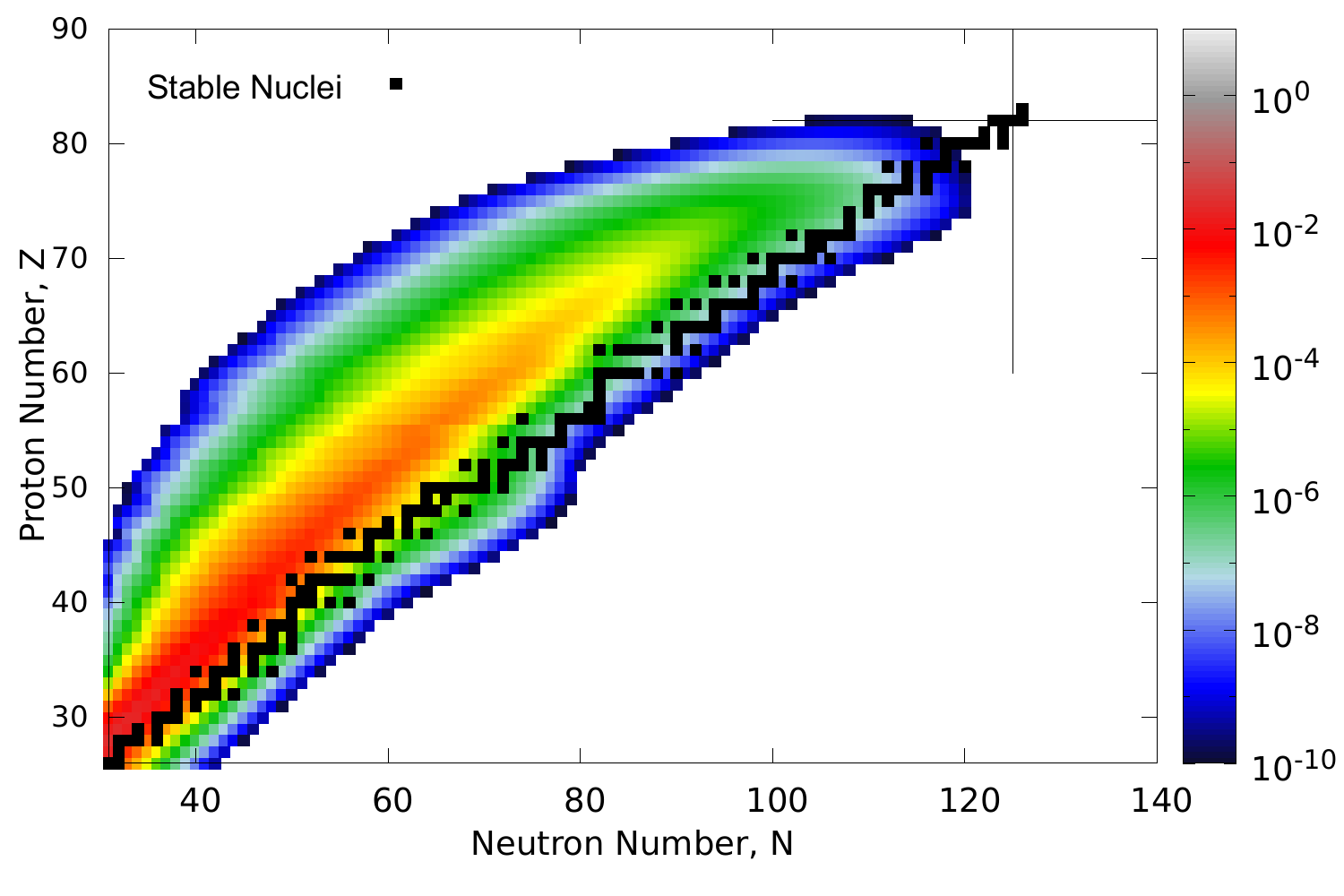}
  \caption{ \label{plots2}  Nuclide charts of total  daughter $A>56 $ isotope yields produced by  the spallation of the QN r-process beam, as a function of proton number (Z) and neutron number (N). The black dots represent stable nuclei. The plots are from top to bottom, $t_{\text{delay}} ({\rm days})=$ 5,3 and 2, respectively. Abundances  were normalized so that the sum of total abundances $A>56$ produced by spallation is unity.}
 
\end{figure}

   
 

 Below we discuss some of the patterns produced by our simulation:
 
\textit{Heavy, r-process elements}:  All isotopes above $A>56$, which are traditionally seen as s-process and r-process yields are abundantly produced.  $A>130$ nuclei produced in QN's r-process spall against $^{56}$Ni and break into $A<130$ nuclei. 
 
\textit{Slope of isotope abundances}: An obvious trend in Fig. \ref{plots1} is how the slope of isotope production is function of time delay. The higher the time delay is, the flatter is the production.  In contrast, a small time delay leads to roughly a negative slope in the plot, where lighter nuclei dominate.  A negative slope is a consequence of a higher $^{56}$Ni target density, which leads to higher amounts of collisions and consequently, the enhancement smaller nuclei. Fig. \ref{plots1} shows how the p-nuclei profile also  "flattens" at high time delays, following roughly a similar trend to the total isotope production.

\textit{Production of proton-rich isotopes}:  Fig. \ref{plots2} shows that most of the spallation products lie in the proton-rich side of the valley of stability. Therefore spallation becomes an extremely efficient site for the production of stable p-nuclei as well. From Fig. \ref{plots2}, we note that the lower the time delay, the more proton-rich is the total distribution of isotopes.

\section{Discussion}

Below we discuss some of the implications of our findings:

\textit{Spallation as a source of $A<130$ r-process isotopes}: Elements in the $90 \le A < 130$ range require a spallation mechanism.  This has some very important consequences. First, this result implies that perhaps, the difficulty of producing $90 < A < 130$ r-process isotopes in either the QN or other, more conventional models like the NSM, is incompleteness - namely, the lack of a spallation mechanism. Therefore, the $90 < A < 130$ solar abundances are actually produced by both spallation and r-process. Spallation therefore, completes other more traditional r-process models by filling the $90 < A < 130$ gap. 

\textit{Spallation as a source of p-nuclei}: $A>130$ r-process isotopes that were produced in the QN ejecta act as "seeds" that are fragmented by spallation into p-nuclei. The spallation reaction highly excites the interacting nuclei, which in turn, de-excite by evaporating mostly neutrons. Thus, spallation by design, tends to produce fragments in the proton-rich side of the valley of stability \citep{cugnon1997nucleon}.  

	Previously, scientists have explored a spallogenic origin for p-nuclei \citep{audouze1970some,hainebach1976cosmic}.  However, the spallation mechanisms previously proposed require high proton fluxes that cannot be reproduced  \citep{woosley1978p}. In contrast, the heavy-ion spallation our model offers is very efficient. A typical QN would eject up to $\sim 10^{-3}$M$_\odot$ of r-process mass, which is almost all spalled in dsQNe with $t_{\text{delay}} \le  5$ days, making the QN's r-process yield an excellent source of p-nuclei seeds. This efficiency leads to a high enhancement of $A < 100$ isotope production, including   $^{92,94}$Mo and $^{96,98}$Ru.  dsQNe with $t_{\text{delay}}\le 5$ days  lead to a  high number of collisions that break the $A>130$ beam into smaller nuclei, which produce the  $A \le 100$ enhancement. 
	
	Our model might offer some clarification on the origin of the rare  $^{138}$La and $^{180m}$Ta. $^{138}$La  can be produced by our spallation model without the need of the $\nu$-process. Finally, It has been argued that high energy spallation can produce the isomer $^{180m}$Ta \citep{hainebach1976cosmic,PhysRevC.26.435,takacs2011activation}.

\textit{Robustness in relation to $^{56}$Ni-synthesis}: The r-process yield is in the order of $\sim 10^{-5}-10^{-3}\text{ M}_{\odot}$, which is much smaller compared to the $\sim 0.1 \text{ M}_{\odot}$ mass of the $^{56}$Ni layer. These large difference in magnitudes implies that the r-process beam will destroy only a trace amount of $^{56}$Ni, leaving the $\sim 0.1 \text{ M}_{\odot}$ figure almost intact. This makes our model, compatible with current light curve measurements  \citep{woosley1991co}, and stellar nucleosynthesis models \citep{woosley1995presupernova}. 

\textit{Robustness in relation to metallicity and stellar evolution}: Studies have shown that r-process abundances in stars are invariant across a  star's metallicity, and therefore it's age, which contradicts current SNII models which are sensitive to a star's initial metal content \citep{sneden2003extremely}. Furthermore, NSMs do not coalesce early in the universe, and thus, cannot explain r-process abundances in very old stars  \citep{jaikumar2007nucleosynthesis}. Moreover, p-nuclei are sensitive to metallicity and other details in a star's evolution, which makes calculating the final, Galactic p-nuclei yield complicated. The dsQN's spallation yield doesn't depend on metallicity nor a star's chemical make-up, which makes it robust against age, metallicity, and stellar mass. Furthermore, dsQN can appear very early in the Universe \citep{ouyed2009quark,ouyed-metal-poor-stars,ouyed2013resolution}. This robustness makes the dsQN an appropriate explanation for the invariant r-process abundances across metallicity. In addition, this robustness makes the dsQN extremely convenient as a source of p-nuclei, for the calculations of the final, Galactic p-nuclei abundances become simplified. 

\textit{Compatibility with r-process mass yield}: Past studies show that the QN rate of event and is mass yield is compatible with the Galactic r-process mass yield \citep{jaikumar2007nucleosynthesis}. It follows, then that p-nuclei mass should be roughly compatible as well.

\textit{Candidates dsQNe}:   The dual-shock QN model 
\citep{ouyed2009predictions} predicted a very specific  super-luminous "double-hump" profile in the dsQN's light curve. Recently, these predictions became true when SN 2009ip and SN 2010mc showed that characteristic "double-hump" and were deemed as dsQNe \citep{ouyed2013sn}. We suggested that CasA was a remnant of dsQN and that it's peculiar chemical nature was a result of spallation  \citep{ouyed2011}. Recently, the dsQN's predictions were hinted as a possible explanation for CasA's unexpected  distribution of Fe and $^{44}$Ti, distributions that were surveyed by NuSTAR \citep{laming2014astrophysics}. CasA was postulated as a $t_{\text{delay}} \sim 5$ days dsQN, which is comparable to the ranges in this paper.   This similarity might point to a degree of universality and/or consistency self-check. Furthermore, if Cas A is a $t_{\text{delay}} \sim 5$ days dsQN, it should show signatures associated with p-nuclei as well. A  $t_{\text{delay}} \sim 5$ days dsQN should produce a few percent of its $A>56$ spallation yield as  p-nuclei. This p-nuclei yield includes the rare $^{138}$La and $^{180m}$Ta isotopes as well. Furthermore, dsQN spallation should produce excited, high angular momentum isomers that would decay and release photons that might be detectable.

\textit{Challenges and future work}: An interesting avenue involves extending our calculations beyond the $^{56}$Ni layer to include the overlaying C and O layers as done in \citep{ouyed2013resolution, ouyed2012}. Furthermore, a more realistic model of the spallation between the QN ejecta and the SN ejecta  would account for the dampening of the beam caused by the gaseous SN ejecta, which would eventually stop spallation and halt the beam in the outer shells.  Due to limitations on the spallation cross section routines, we couldn't include isotopes much heavier than $A\sim 208$, which requires a more sophisticated spallation algorithm. Furthermore, the spallation algorithm we used does not discriminate between isomers, so the production of $^{180m}$Ta isn't calculated quantitatively.  This simulation is very useful for pointing at broad, qualitative patterns, but further research requires a more sophisticated code, with hydrodynamics and more detailed nuclear physics included. Another task needs to be done is to make a statistical comparison with solar abundances through a galactic chemical evolution process.

\textit{Acknowledgments}: We thank Z. Shand for assistance in plotting Fig. \ref{plots2}.  This research is supported by an operating grant from the National Science and Engineering Research Council of Canada (NSERC). 

\bibliographystyle{elsarticle-num} 
\bibliography{ouyed2012}

\end{document}